\documentclass[12pt]{article}
\usepackage{epsfig}

\newcommand{\lapprox}{\lower.6ex\hbox{$\; \buildrel<\over\sim \;$}}

\newcommand{\gapprox}{\lower.6ex\hbox{$\; \buildrel>\over\sim \;$}}

\newcommand{\curly}{\lower1.ex\hbox{$\; \stackrel{\textstyle \wr}{\wr} \;$}}

\def\barr{\begin{array}}
\def\earr{\end{array}}
\def\berr{\begin{eqnarray}}
\def\err{\end{eqnarray}}
\def\berrno{\begin{eqnarray*}}
\def\errno{\end{eqnarray*}}
\def\be{\begin{equation}}
\def\ee{\end{equation}}

\def\fr{\frac}
\def\noin{\noindent}

\def\apj{{\it Astrophys.~J.}}

\def\prd{{\it Phys.~Rev.~D.}}

\def\AnA{{\it Astr. Astrophys.}}
\def\grg{{\it Gen. Rel. Grav.}}
\def\plb{{\it Phys. Lett. B}}

\renewcommand{\a}{\alpha}

\renewcommand{\t}{\theta}

\title{ \bf Power law cosmology - a viable alternative }
\author{
{Abha Dev} \\
{Miranda House, University of Delhi, Delhi 110 007}\\
\\
{Deepak Jain {\footnote{djain@ddu.du.ac.in}}} \\
       { Deen Dayal Upadhyaya College, University of Delhi, New Delhi 110 015 } \\
{and}\\
{Daksh Lohiya {\footnote{dlohiya@gmail.com}}}\\
{ Department of Physics and Astrophysics,} \\
       { University of Delhi, Delhi-110007, India}\\
}
\begin{document}

\maketitle
 
\begin{abstract}
\noindent A power law cosmology  is defined by the cosmological
 scale factor evolving  
as $t^\alpha$. In this work, we put bounds on  
$ \alpha$ by  
 using  the joint test of the SNe Ia data from Supernova  Legacy Survey (SNLS)
and $H(z)$ data with curvature constant $ k = 0, \,\pm 1$. We observe that
the combined analysis with SNLS and H(z) data favours the open power
 law cosmology with $ \alpha = 1.31^{+0.06}_{-0.05}$.
It is also interesting to note that an Einstein - de
 Sitter model ($\alpha = 2/3$) is  ruled out 
 at $ 2 \sigma $ level.

\end{abstract}

\section{Introduction}

\noin Our universe is very well explained by the Standard Cosmological Model (SCM)
based on  the Hot Big-Bang theory  and the  inflationary scenario. However, 
there are still some features of the universe which cannot
be understood within the SCM. One of the major unsolved problem is
the {\it cosmological constant problem}. The standard model fails to explain
why the energy density of the vacuum is 120 orders of magnitude smaller
than its value at the Planck time \cite{sami}. The problems in the SCM and
the availability of precise data from various  observations 
have encouraged cosmologists to explain the observed universe 
through alternative cosmological models. 

\vskip 0.3cm
\noindent One of the interesting alternatives is a
{\bf Power Law Cosmology}. In such a  model there is a
power law evolution of the cosmological scale factor, $a(t)
 \propto t^\alpha$. The power law evolution with $\alpha \ge 1$ has
 been discussed at length in a series of earlier articles 
\cite{kolb,mann,allen}. The motivation for such a scenario comes from
the fact that it does not encounter
 flatness and the horizon problem at all. Another interesting
 feature of these models is that they easily accommodate high
 redshift objects and hence alleviate the age problem. 
 These models are also purged of the fine tuning problem
\cite{dol,ford}. Such a scaling is a generic feature in a class of
 models that attempt to dynamically solve the cosmological constant problem
 \cite{mann,allen,dol,ford,wein}.

\vskip 0.3cm
\noindent   A power law evolution of the cosmological scale factor 
with $\alpha \approx 1$ is surprisingly an excellent fit to a 
host of cosmological observations. Any model that can support such 
a coasting presents itself as a falsifiable model as far as classical cosmological 
tests are concerned as it exhibits distinguishable and verifiable 
features. An evolution of this nature is supported by classical 
cosmological tests such as the galaxy number counts 
as a function of redshift  and the data on angular diameter 
distance as a function of redshift \cite{kolb}. However, as these tests are 
marred by evolutionary effects (e.g. mergers), they have fallen 
into disfavour as reliable tests of a viable model. 
With the discovery of Supernovae type Ia, SNe Ia, as reliable standard candles, the 
status of the Hubble test has been elevated to that of a precision
measurement. The Hubble plot relates the magnitude of a standard 
candle to its redshift in an expanding FRW universe. We demonstrated
that linear coasting cosmology 
accommodates the high redshift objects while the standard model could
not \cite{abha,sethi}. Such a model is also comfortably consistent with the
gravitational lensing statistics 
\cite{abha} and the primordial nucleosynthesis \cite {annu}.

\vskip 0.3cm
\noin  The plan of the paper is as follows. In Section 2, we
give the basic equations for open, closed and flat power law scenarios.  
In Section 3, we  
find the constraints on the cosmological parameter $\alpha$ from a joint
test of the Supernova Legacy Survey SNe Ia data set (SNLS) and the
H(z) data. The joint test is performed for open, closed and
flat power law cosmologies. The results are summarized in Section 4.

\section{Power Law Cosmology}

\noin For a FRW metric, the line element is 
\be
ds^2=c^2dt^2-a^2(t) \left[\frac{dr^2}{1 -k r^2} + r^2(d\t^2
+\sin^2{\t}\,d\phi^2)
\right]\,\,.
\label{eq:ds}
\end{equation}

\noin Here $k =\pm1,0$, is the curvature constant,
$t$ is the cosmic proper time and $a(t)$ is the
cosmological scale factor. 

\vskip 0.3cm
\noin The expansion rate of the universe is described by a Hubble
parameter, $H(t) = 
\dot{a}/a$. The present expansion rate of the universe is
defined by the {\it Hubble constant} $H_0$. Here
and subsequently the subscript 0 on a parameter refers to its present value.

\vskip 0.3cm
\noin In this paper, we study a general power law cosmology with the
scale factor given in 
terms of a dimensionless parameter $\a$
\be
a(t)= \frac{c}{H_{0}}\left(\fr{t}{t_0}\right)^{\a}\,\,.
\label{eq:a}
\end{equation}

\vskip 0.3cm
\noin In this  model, $H(t) = \alpha/t$ and $H_0=\a/t_0$.
\noin The scale factor and the redshift at time $t$ are related to
their present values by 

\be
\frac{a_0}{a(z)} = \frac{t_0}{t} = 1+z\,\,.
\label{eq:ao/a}
\end{equation}

\noin The present `radius' of the universe is defined as
\be
a_0 =\frac{c}{H_{0}}\,\,.
\label{eq:a_0}
\end{equation}

\noin The age of the universe at redshift $z$
is given as 
\be
t(z)=\fr{\a}{H_0(1+z)^{1/\a}}\,\,.
\label{eq:age}
\end{equation}
\noin The dimensionless Hubble parameter is defined as:  
\be
E(z) \equiv \frac{H(z)}{H_0}=(1+z)^{1/\a}.
\label{eq:dimensionless_hubble}
\end{equation}

\vskip 0.3cm
\noin For the power law cosmology, the luminosity distance 
between two redshifts $z_{1}$ and $z_{2}$ is 
\be
d_{\rm L}(z_{1},z_{2}) = \fr{c(1+z_{2})}{H_{0}} \,\,
\mathcal{S} \left(\fr{\a}{\a-1}\left
  \{(1+z_{2})^{\fr{\a-1}{\a}}-(1+z_{1})^{\fr{\a-1}{\a}}\right \}
  \right)\,\,. 
\label{DL}
\end{equation} 

\vskip 0.3cm
\noindent In the limiting case, $\a \rightarrow 1$, we obtain 
\be
d_{\rm L}(z_{1},z_{2}) = \fr{c\,\,(1+z_{2})}{H_{0}}\mathcal{S}
\left[ ln(1+z_2)-ln(1+z_1)\right] \,\,.
\end{equation}

\vskip 0.3cm

\noindent Here

\berr
\mathcal{S}(x) &=& \sinh(x)\,\,\mathrm{for}\,\,\, k = -1 \nonumber\\
&=& \sin(x)\,\,\mathrm{for} \,\,\,k = +1 \nonumber\\
&=& x \,\,\mathrm{for}\,\,\, k = 0 \nonumber\\
\err

\section{ Observational Tests}

\subsection{Constraints from the Supernova Legacy Survey SNe Ia data
set (SNLS)}

\noin Type Ia supernova (SNe Ia) are excellent
cosmological standard candles for estimating the apparent magnitude
$m(z)$ at peak brightness after accounting for  various corrections. 
In this work we use the SNLS data set of 115 SNe Ia data points with
redshift $ z< 1$ \cite{snls}.

\vskip 0.3cm
\noin For a standard candle  of absolute magnitude M, the apparent
magnitude $m(z)$ can be expressed as:

\be
m(z) = {\mathcal M} + 5 \,\log_{10}D_L(z)\,\,.
\ee

\noin Here $D_L(z)$ is related to the luminosity distance:
{\setlength\arraycolsep{2pt}
\berr
D_L(z) & = & \fr{H_{0}}{c} d_{\rm L}(0,z) \nonumber\\
& = & (1+z)\,\,\mathcal{S}\left[\fr{\a}{\a-1}\left
  \{(1+z)^{\fr{\a-1}{\a}}-1\right \}
 \right]\,\, 
\err}

\noin and
{\setlength\arraycolsep{2pt}
\berr
{\mathcal M}& =& M - 5 \log_{10}h + 42.38 \,\,,
\err}

\vskip 0.3cm
\noindent is the ``zero point'' magnitude. We use $H_0 \,= \, h \,100
\,\mathrm {Kms^{-1}\, Mpc^{-1}}$.
\\
\\
\noindent The  distance modulus, $\mu(z)$, is defined as
\be
\mu (z) = m(z) - M=  5 \log_{10}D_L(z) - 5 \log_{10}h + 42.38. 
\ee

\vskip 0.3cm
\noindent The  {\it chi-square} function is defined as
\be
\chi^{2}(h, \alpha) =
\sum_{i=1}^{115} \left [ \frac{{\mu_{exp}^{i}(h,\alpha,
    z_i)-\mu_{obs}^ i}(z_i)}
{\sigma_i} \right ]^{2},
\label{eq:SNe}
\ee

\noindent where $\mu_{exp}$ is  the expected distance modulus for a
supernova at a given redshift z and $\sigma_i$ is the
error due to intrinsic 
dispersion of SNe Ia absolute magnitude and observational 
uncertainties in SNe Ia peak luminosity. These errors are assumed to
be Gaussian and uncorrelated. The observed distance
modulus, $\mu_{obs}$,  is given by the supernovae data set.

\subsection{Constraints from H(z) data}

\noin Simon, Verde and Jimenez (2005) used differential ages of passively
evolving galaxies to determine the Hubble parameter as a function of
redshift, $H(z)$ \cite{svj}. They use a sample of absolute
ages of 32 
galaxies taken from the Gemini Deep Deep Survey (GDDS) and the archival
data to obtain 9 data points of $H(z)$ with  $0.09 \le z \le 1.75$.  
The Hubble parameter and the differential age of the universe, dz/dt,
are linked by the equation: 
\be
H(z) = -{1 \over {1 +z}}{dz \over dt}.
\end{equation} 
The details of the method for calculating the dz/dt from the absolute
age is given by Simon, Verde and Jimenez (2005) \cite{svj}.

\vskip 0.3cm
\noindent The $H(z)$ for power law cosmology is given by: 
$$
E(z) \equiv \frac{H(z)}{H_0}=(1+z)^{1/\a}.
$$
\noindent In order to put bounds on the model parameter, $\alpha$,
we define the quantity:
 \be
\chi^2 (h,\alpha)= \,\sum_{i=1}^9 { \left ( H_{exp}(z_i,\alpha,h) -
  H_{obs}(z_i) \right)^2 \over {\sigma_i^2}}
\label{eq:Hz}
\ee

\vskip 0.3cm
\noindent Where $H_{exp}$ is the expected value of the Hubble constant
in the power law cosmology, $H_{obs}$ is the observed value and
$\sigma_i$ is 
the corresponding $1\,\sigma$ uncertainty in the
measurement.

\subsection{Joint Test: SNe Ia + H(z)}

\noin We find the constraints on the cosmological
parameter $\alpha$ from the joint test of SNe Ia and H(z) data sets.
In this joint test we define the quantity

\be
\chi^2_{\mathrm{joint}} = \chi^2_{\mathrm{SNe}} + \chi^2_{\mathrm{H(z)}},
\ee

\noin where $\chi^2_{SNe}$ is given by Eq.(\ref{eq:SNe}) and
$\chi^2_{H(z)}$ by Eq.(\ref{eq:Hz}).

\vskip 0.3 cm
\noin Considering $h$ to be a nuisance parameter, we
marginalize over $h$ to obtain the
probability distribution function defined as:

$$ L( \alpha) = \int \, e^{-\chi^2_{\mathrm{joint}}(h,\,\,\alpha)/2}\,P(h)\,\,
dh.$$

\vskip 0.3cm
\noindent Here $P(h)$ is the prior probability  function for $h$ which
we assume to be Gaussian:  

$$P(h) = {1 \over {\sqrt{2\pi} \sigma_{h}}} \,\,\exp[-\frac{1}{2} \,\,\frac{(h-
h_{obs})^2}{\sigma_{h}^{2}}],
$$

\vskip 0.3cm
\noin  where $h_{obs}$ is the value of $h$ (and $\sigma_h$ is the error
in it) as suggested by independent observations. In this paper, we
also study the effect of different priors on the result. We use two
set of priors: 
\vskip 0.3cm

\noin (i) {\bf Set A:} $ h_{obs}\,\,=\, 0.68 \pm 0.04$ as obtained from
the median statistics analysis of 461 
measurements of $H_0$ \cite{chen}. 

\vskip 0.3cm
\noin (ii) {\bf Set B:} $h_{obs}\,\, =\,0.77 \pm 0.04$ as
suggested by the Chandra  X - ray  Observatory results \cite{cxo}.  

\vskip 0.3cm
\noin The best fit model parameter is obtained by minimizing the
modified $\chi^2$ (obtained after marginalization over h) :

\be
\chi^2 = -2 \ln L(\alpha)
\ee

\noin We performed the joint analysis for open, closed and
flat power law cosmologies. 
The best fit value of $\alpha$ and the constraints on it seem to be
independent of the choice of prior in all the three models. The
result obtained with both the priors are summarised in Table 1. The
joint analysis also shows that the best fit scenario is an open model with
$\a_{min} = 1.31^{+0.06}_{-0.05}$.

\begin{table}

\begin{center}
\begin{tabular}{l l r}\hline\hline
{\bf Closed} & {\bf Flat} & {\bf Open} \\          
\hline
\hline
&& \\
{\bf Set A Prior}&&\\
&& \\
$\chi^2_{\nu}= 1.51$ & $\chi^2_{\nu}= 1.28$ & $\chi^2_{\nu}=1.15$\\
& & \\
$\a_{min} = 2.28^{+0.23}_{-0.19}$ &
$\a_{min} = 1.62^{+0.10}_{-0.09}$ & $\a_{min} = 1.31^{+0.06}_{-0.05}$ \\ 
& & \\
&&\\
{\bf Set B Prior} & &\\
&& \\
$\chi^2_{\nu}= 1.53$ & $\chi^2_{\nu}= 1.30$ & $\chi^2_{\nu}=1.17$ \\
& & \\
$\a_{min} = 2.28^{+0.23}_{-0.19}$ &
$\a_{min} = 1.62^{+0.10}_{-0.09}$ & $\a_{min} = 1.31^{+0.06}_{-0.05}$\\ 
& & \\
\hline 
\end{tabular}
\caption{The best fit value of $\a$  and $\chi^2_{\nu}
  = \chi^2_{min} /( \mathrm {degree\,\, of \,\, freedom})$ for
the three models.}

\end{center}

\end{table}

\section{Summary}

\noin In this paper, we study observational constraints on the power law
cosmology, $a(t) \, \propto t^{\alpha}$. This model of the universe
has very interesting features which makes it unique when compared to
the other models of the universe. Firstly, for  $ \alpha \ge 1$ it
does not encounter the horizon, flatness and age 
problem \cite{kolb, mann, allen}. Secondly, such an evolution is a
 characteristic feature of
models that dynamically solve the 
cosmological constant problem. Statistically this model may be
preferred over other models as we have to fit only one parameter,
$\alpha$. 

\vskip 0.3cm
\noin  In the work presented here, we use the joint test, which uses
the SNLS data and H(z) data, to put constraints on the parameter
$\alpha$. To begin with, we work with all the three models - closed, 
flat and open and put constraints on $\alpha$ in these three cosmologies.
The results on the cosmological test are summarized in Table 1. Fig. 1-3
show variation of $\chi^2$ with $\alpha$ for the three models ( with
set B prior). We also mark the 
parametric space allowed at 90\% CL in the figures.
We make the following observations:

\noindent $\bullet$ For the three models under consideration,
the value of $\chi^{2}_{\nu}$, the best fit value of $\alpha$ 
and the constraints on it seem to be independent of the choice of prior.
\vskip 0.2cm
\noin $\bullet$ The joint test favours an open power law cosmology 
with $\a_{min} = 1.31^{+0.06}_{-0.05}$. As can be seen in Table 1, $\chi^2_{\nu}$
is minimum for an open power law cosmology. 

\vskip 0.2cm
\noin$\bullet$ The joint analysis does not rule out flat and closed power law  cosmologies. 
However, we do observe that the 
constraints on $\alpha$ are tighter for an open model.  

\vskip 0.2cm
\noin $\bullet$ The joint analysis rules out linear coasting ($\alpha = 1$)
in all the three cosmologies even at  90 \% CL. (see Fig.1, Fig.2 and Fig.3).

\vskip 0.3cm
\noin These observations match the conclusions 
of Zhu et al. (2007) \cite{zhu}. They test the  power-law cosmology 
against the recent measurements of the X-ray gas mass fractions 
in clusters of galaxies. They conclude  that
the best fit is an  open model (with $\a_{min} = 1.14 \pm 0.05$) 
though the flat and closed models can not be ruled out.

\vskip 0.3cm
\noindent In the past, various observational tests have been used to put
constraints on the parameter $\alpha$ in an open power law cosmology,
such as gravitational 
lensing, Old High Redshift Galaxies (OHRG), SNe Ia and X-ray gas mass
fractions in galaxy
clusters. Constraints obtained from the other tests along with the 
constraints obtained from the SNe Ia data and H(z) data are summarized
in Table 2. The interest in an open power law cosmology 
is on  account of the fact that a whole class of dilaton
gravity models that dynamically solve the cosmological constant rely
on a vanishingly small effective gravitational constant in the early
universe \cite{wein, meetu}. This gives an open FRW model for any
reasonable equation of state of matter.

\vskip 0.3cm
\noin We observe that the joint analysis of SNLS and 
H(z) data favours an open power law cosmology with $\alpha > 1$.
We further observe that the joint  analysis (SNLS + H(z)) done in 
this letter rules out the Einstein-de Sitter universe ($\a = 2/3$). 
\vskip 0.3cm
\noin
Since the joint test of SNLS and 
H(z) data presented in this work favours an open 
power law cosmology, for the sake of completeness we find bounds 
on $\alpha$ in open model separately using the SNLS data and the H(z) data. 
We once again marginalize over $h$ to 
find $\chi^2_{\nu}$ and the best fit values using each test. We find that:

\vskip 0.2cm
\noindent {\bf 1.} {\bf Constraints from SNLS data:} For set A prior, we get
the best fit value $\alpha = 1.421_{-0.07}^{+0.08}$ 
with  $\chi^2_{\nu} =  1.07$ 
With  set B, we get the same constraints on the parameter
$\alpha$ as obtained from set A but  with  $\chi^2_{\nu} =
1.09$. We find that the constraints on $\alpha$ do not depend upon the
choice of the prior. 
We, therefore, conclude that the SNLS data favours $\alpha > 1$ 
(best fit value being $\alpha = 1.42^{+0.08}_{-0.07}$ ). This observational data rules out
linear coasting cosmology ($ \alpha = 1$) even at at $2 \sigma$
level. 

\vskip 0.2cm
\noindent {\bf 2.} {\bf Constraints from H(z) data:} We find that the 
H(z) data provides tight constraints on the model parameter $\alpha$. 
With set A prior we obtain best fit value 
$\alpha = 1.02_{-0.06}^{+0.09}$ with  $\chi^2_{\nu} =  0.834$.
With set B prior we get  the best fit value as
$\alpha = 1.07_{-0.09}^{+0.11}$ with  $\chi^2_{\nu} =
1.06$. We observe that for this test the constraints on $\alpha$
weakly depend upon the choice of priors. The H(z) data, however, 
strongly favours linear coasting cosmology with the best fit value.

\vskip 0.2cm
\noindent We summarize: An open  power law cosmology with $\alpha >
1$ is in excellent agreement with the present day observations. 
This makes it an attractive alternative. In fact, the possibility 
of an open  linear coasting model as a viable model cannot be ruled
out (as suggested by the H(z) data). Concordance of the power-law cosmology 
with CMB anisotropy measurement is a major area to be explored. 
There are large numbers of surveys that are ongoing or have been
proposed. With the flood of new data (and the possibility that the
observational  
techniques will be improved), the task ahead is to find  models
of the universe that can explain these observations. It will be
interesting to investigate how the future observations will change the
constraints on $\alpha$.

\begin{table}
\begin{center}
\begin{tabular}{l l r}\hline\hline
Method &  Reference & $\a$  \\          
\hline
\hline
Lensing Statistics &  & \\
& & \\
(i) $n_{\rm L}$ & Dev et al.\,\cite{abha} &  $1.09 \pm 0.3$ \\
(ii) Likelihood &Dev et al.\,\cite{abha}  & $1.13^{+0.4}_{-0.3}$\\
\hspace{0.5cm} Analysis  &  &   \\
& & \\
OHRG & Dev et al.\,\cite{abha} & $\ge 0.8$ \\
& & \\
SNe Ia (Gold sample)  & Sethi et. al \,\cite{sethi} & $1.001$ \\
& & \\
Old Quasar & Sethi et. al.\,\cite{sethi} & $\ge 0.85$\\
& & \\
Galaxy Clusters & Zhu et. al.\,\cite{zhu} &$ 1.14 \pm 0.05$ \\
&&\\
SNe Ia (SNLS)& This Letter & $1.42^{+0.08}_{-0.07}$\\
&&\\
H(z) data & This Letter & $1.07^{+0.11}_{-0.09}$ \\
&& \\
SNe Ia + H(z) data & This Letter & $1.31^{+0.06}_{-0.05}$ \\
&& \\
\hline 
\end{tabular}
\caption{Constraints on $\a$ from various cosmological tests in an open
  power law cosmology model.}
\label{T2}
\end{center}
\end{table}

\section*{Acknowledgments}

\noin A. Dev $\&$ D. Jain thank Amitabha
 Mukherjee and Shobhit Mahajan for providing  the facilities to carry out the research
 work.

\begin{figure}[ht]
\vspace{-1.2in}
\centerline{
\epsfig{figure=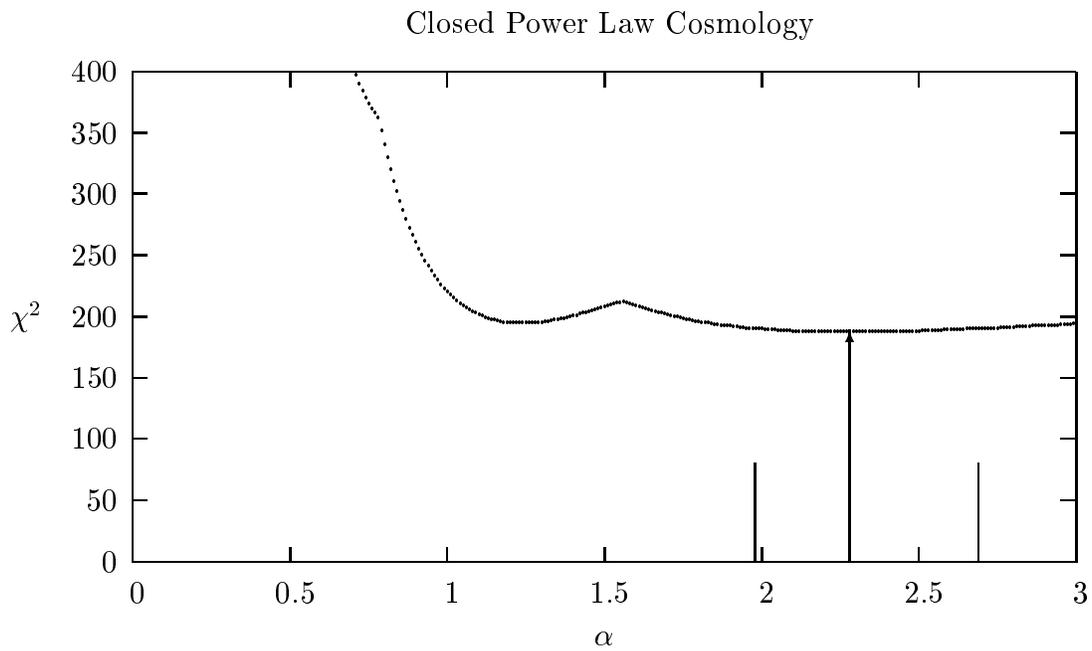,width=1.5\textwidth}}
\vspace{-5.2in} 
\caption{Results with Gaussian prior $h_{obs}= 0.77 \pm 0.04$ in a closed power 
law cosmology. The vertical lines at $\alpha = 1.98$ and at $\alpha = 2.69$ mark the
parametric space allowed at 90\% CL. The minimum of $\chi^2$
occurs at $\alpha = 2.28$ .}
\end{figure}

\vfill
\eject

\begin{figure}[ht]
\vspace{-1.2in}
\centerline{
\epsfig{figure=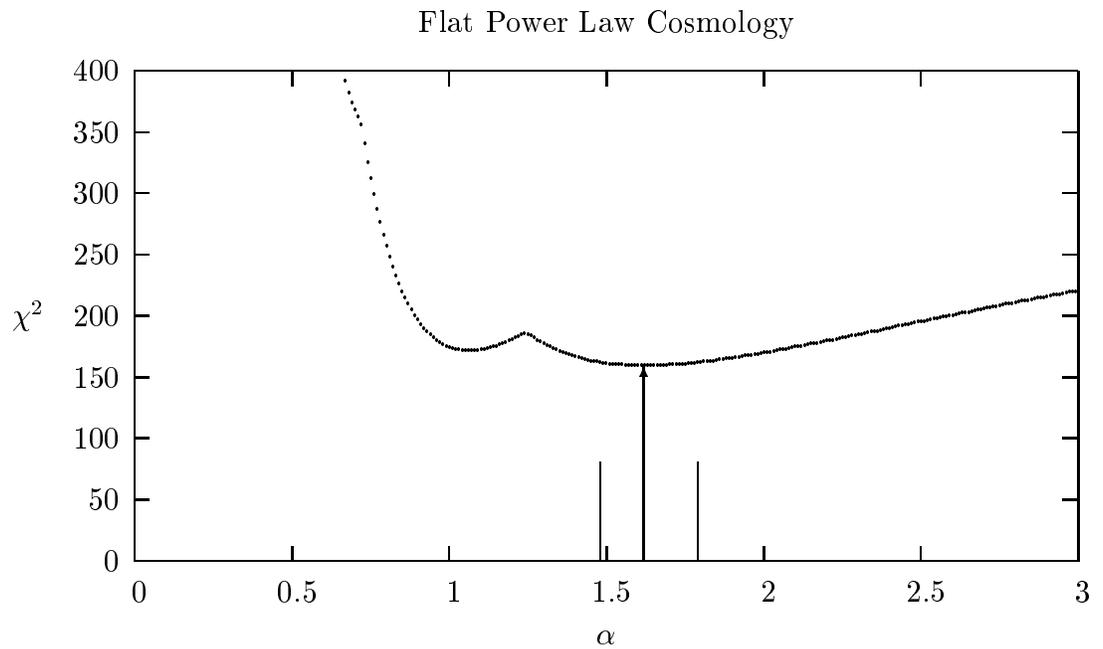,width=1.5\textwidth}}
\vspace{-5.2in} 
\caption{Results with Gaussian prior $h_{obs}= 0.77 \pm 0.04$ in a flat model. The
  vertical lines at $\alpha = 1.48$ and at $\alpha = 1.79$ mark the
parametric space allowed at 90\% CL. The best fit value occurs at $\alpha =
1.62$. }
\end{figure}

\vfill
\eject

\begin{figure}[ht]
\vspace{-1.2in}
\centerline{
\epsfig{figure=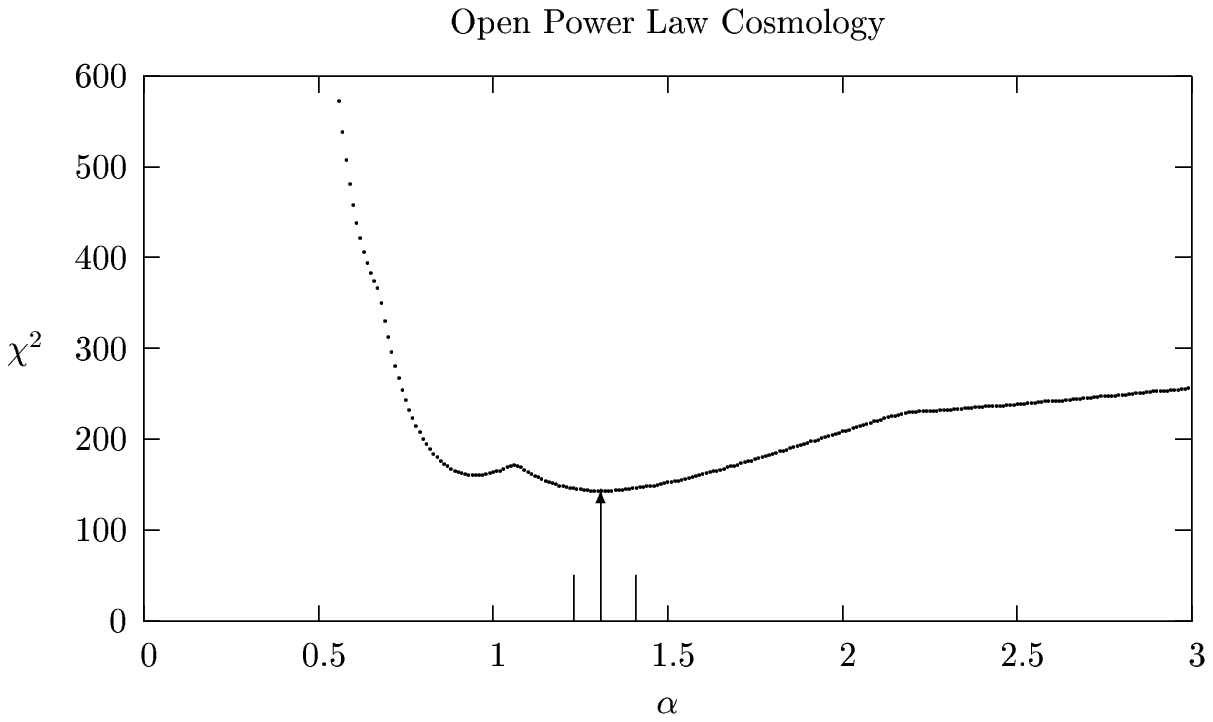,width=1.5\textwidth}}
\vspace{-5.2in} 
\caption{Results with Gaussian prior $h_{obs}= 0.77 \pm 0.04$ in an open
model. The 
vertical lines at $\alpha = 1.23$ and at $\alpha = 1.41$ mark the
parametric space allowed at 90\% CL. The minimum occurs at $\alpha =
1.31$ .}
\end{figure}


\begin{thebibliography}{10}

\bibitem{sami} Copeland, E. J., Sami, M. and  Tsujikawa, S. (2006),
  {\it Int. J. Mod. Phys.}, {\bf D15}, 753.

\bibitem{kolb} Kolb, E. W. (1989),\apj, {\bf 344}, 543.

\bibitem{mann} Manheim, P., and Kazanas, D. (1990), \grg, {\bf 22}, 289.

\bibitem{allen}Allen, R.~E. (1999),  arXiv: astro-ph/9902042.

\bibitem{dol}Dolgov, A.~D. (1982) in the {\it The Very Early Universe},
 eds. Gibbons, G., Siklos, S., Hawking, S.~W. (Eds.),  Cambridge
 ~Univ. Press ; Dolgov, A.~D. (1997), \prd,{\bf 55}, 5881. 

\bibitem{ford} Ford, L.~H. (1987), \prd, {\bf 35}, 2339. 

\bibitem{wein} Weinberg, S. ( 1989), {\it Rev. Mod. Phys.}, {\bf 61}, 1.

\bibitem{abha} Dev, A. et al. (2002), {\plb}, {\bf 548}, 12.

\bibitem{sethi} Sethi, G., Dev, A. and Jain, D. ( 2005), \plb, {\bf
  624}, 135.

\bibitem{annu} Batra, A. et al. (2000),
 {\it Int. J. Mod. Phys.}, {\bf D9}, 757; Batra, A., Sethi, M.,
  and  Lohiya, D.  (1999), \prd, {\bf 60}, 108301. 

\bibitem{sav} Gehlaut, S. et al. (2002), {\it Spacetime Substance},
  {\bf 4}, 14.

\bibitem{snls} Astier, P. et al. (2006), \AnA, {\bf 447}, 31. 

\bibitem{svj} Simon, J., Verde, L. and Jimenez, R. ( 2005), \prd, {\bf
  71}, 123001.

\bibitem{chen} Chen, G. et al. (2003), PASP, {\bf 115}, 1269.
 
\bibitem{cxo} Bonamente, M. et al. (2005), arXiv: astro-ph/051349.


\bibitem{zhu}Zhu, Z. H. et al. (2007), arXiv: astro-ph/0712.3602.

\bibitem{meetu} Lohiya, D. and Sethi., M (1999), {\it Class. Quant.
  Grav.}, {\bf 16}, 1545. 

\end{thebibliography}
\end{document}